# The Kinetic Basis of Morphogenesis

Yuri Shalygo[1]

[1]Gamma Ltd, Vyborg, Russia
yuri.shalygo@gmail.com

## Abstract

It has been shown recently (Shalygo, 2014) that stationary and dynamic patterns can arise in the proposed one-component model of the analog (continuous state) kinetic automaton, or kinon for short, defined as a reflexive dynamical system with active transport. This paper presents extensions of the model, which increase further its complexity and tunability, and shows that the extended kinon model can produce spatio-temporal patterns pertaining not only to pattern formation but also to morphogenesis in real physical and biological systems. The possible applicability of the model to morphogenetic engineering and swarm robotics is also discussed.

## Introduction

In his seminal paper on morphogenesis (Turing, 1952), Alan Turing demonstrated that different spatio-temporal patterns can arise due to instability of the homogeneous state in reaction-diffusion systems. It has been shown recently (Shalygo, 2014) that stationary and dynamic patterns can also arise in the proposed one-component model of the analog (continuous state) kinetic automaton, or kinon for short, defined as a reflexive dynamical system with active transport. This paper presents extensions of this model, increasing further its complexity and tunability. The main aim of this paper is to show that anisotropic diffusion, usually regarded as anomalous, in fact is quite ubiquitous and can be harnessed in morphogenetic engineering and robotics.

The proposed model stems from a number of existing models of complex dynamical systems, and the following in particular:

- *Cellular Automata* (CA) conceived in 1950's by John von Neumann and Stanislaw Ulam.
- *Coupled Map Lattices* (CML) proposed in 1985 by Kunihiko Kaneko as a paradigm for the study of spatio-temporal complexity.
- *Lattice Gas Automata* (LGA) introduced in 1986 by Frisch et al and Stephen Wolfram independently for the modeling of fluid dynamics.
- *Lattice Boltzmann Model* (LBM) evolved from LGA and attracting growing popularity in Computational Fluid Dynamics (CFD) and other fields (Chopard et al, 2002).

Nevertheless, a decisive impetus for the kinon model was given by Konrad Zuse's *net automaton* (Zuse, 1969) and Gordon Pask's *diffusion network* (Pask, 1961) [Fig.1].

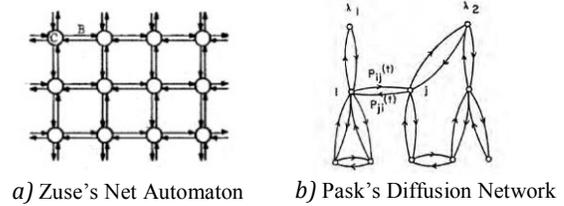

*a)* Zuse's Net Automaton  *b)* Pask's Diffusion Network

Fig.1 *Dynamic systems with active transport*

The central idea behind these networks is that nodes of the network are connected reciprocally with the lines that have not only transport but also storage functions. This is in sharp contrast to the conventional view on network links as passive elements. None of the existing models can be applied to Pask's diffusion networks; therefore, a new generation of topology and state space invariant models with active links is needed. The basic kinon model, introduced in the previous paper and outlined in the next paragraph, is a trial step in this direction.

## Background

The majority of the existing models are discrete time networks, in which values assigned to nodes and representing their current state are updated synchronously by some transformation. According to the type of transformation, they can be divided in two main groups: *functional* and *relational*.

*Functional transformation* maps a set of input values onto a single (scalar) output value - a new state of the node, which is relayed or fanned out to all output links. Formally, functional transformation is a many-to-one map $F: Q^{k+1} \to Q$, where $Q$ is a set of states of the node and its $k$ neighbors [Fig.2a].

*Relational transformation* maps a set of input values onto a set (vector) of output values of the same dimension, therefore it is homomorphism or a structure preserving (one-to-one) map of the form $R: Q^{k+1} \to Q^{k+1}$ [Fig.2b].

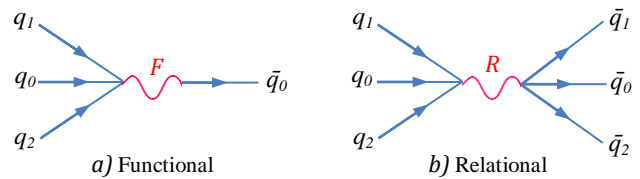

*a)* Functional  *b)* Relational

Fig.2 *Feynman diagrams of state transformation*

The difference stems from different model structures. In functional models, the state of a cell is represented by a scalar value $q_0$ [Fig.3a]. In relational models, it is a vector $\{q_0, q_1...q_k\}$ associating the first component with a cell and the other ones with its $k$ neighbors [Fig.3b]. Contrary to functional models, the value $q_0$ is not observable to the neighbors of the cell. The values $q_1...q_k$ represent the feedback (observables) of the cell and reside in the links responsible both for information propagation and storage. It implies the *dualism* of relational models, reincarnating as *autonomous cells* during collision or *autonomous links* during propagation.

The model represented in Fig.4 was elaborated with having in mind Rosen's Modeling Relation (Rosen, 1991) as well as Kauffman's autonomous agent doing its own work-constraint cycle (Kauffman, 2000).

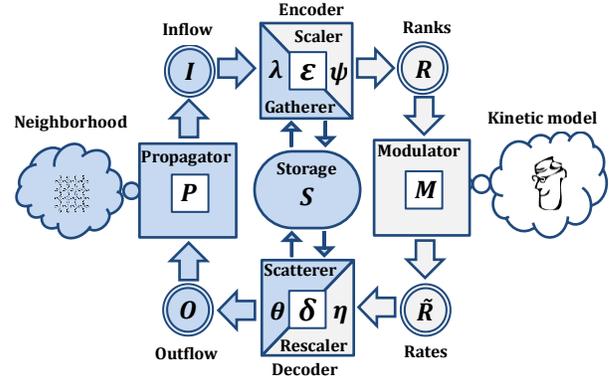

Fig.4 *Kinon State-Transition Structure*

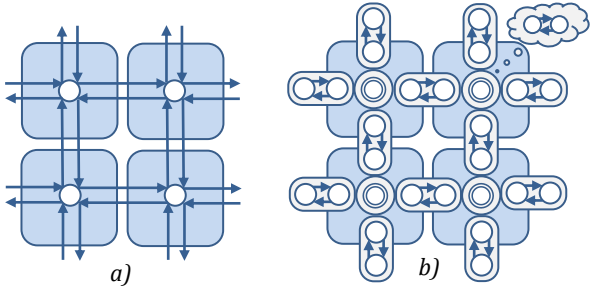

Fig.3 *Structure of functional (a) and relational (b) models*

Similar to LGA and LBM, the kinon model is *relational* and *quantity conservative*, because it was designed to be able to simulate real physical phenomena. CA, CML, Random Boolean Networks, Artificial Neural Networks, etc. are functional and non-conservative in general.

The kinetic automaton can be viewed as the generalization of LBM, which is not restricted to the Boltzmann equation and a regular grid. The key element of the model, making its properties and dynamics different from LBM, is a collision step which was transformed into a 3-step operator (Encoding-Modulation-Decoding) called Conservative Rank Transform (CRT). In this method, not quantities as such but their relative values (ranks) are transformed (modulated), and the total quantity does not change after transformation.

Formally, a kinon is a 9-tuple $(I, R, \tilde{R}, O, S, P, \varepsilon, M, \delta)$, where:

$I$ - vector of absolute $\mathfrak{R}^+$ input values $(I_1... I_k)$ *(inflow)*,
$R$ - vector of relative *[0,1]* input values $(R_0, R_1...R_k)$ *(ranks)*,
$\tilde{R}$ - vector of relative *[0,1]* output values $(\tilde{R}_o, \tilde{R}_1... \tilde{R}_k)$ *(rates)*,
$O$ - vector of absolute $\mathfrak{R}^+$ output values $(O_1...O_k)$ *(outflow)*,
$S$ - two-vector of absolute $\mathfrak{R}^+$ values $(S_i, S_o)$ *(storage)*,
$P$ - propagation operator $P: O \to I$ *(propagator)*,
$\varepsilon$ - encoding operator $\varepsilon: (I, S_o) \to (R, S_i)$ *(encoder)*,
$M$ - modulation operator $M: R \to \tilde{R}$ *(modulator)*,
$\delta$ - decoding operator $\delta: (\tilde{R}, S_i) \to (O, S_o)$ *(decoder)*.

Encoding ($\varepsilon$) is a composition of gathering ($\lambda$) and scaling ($\psi$) operators: $\varepsilon = \psi \circ \lambda$. Similarly, decoding ($\delta$) is composed of rescaling ($\eta$) and scattering ($\theta$) operators: $\delta = \theta \circ \eta$.

Schematically, the kinon internal structure can be represented in more detail by the diagram in Fig.5:

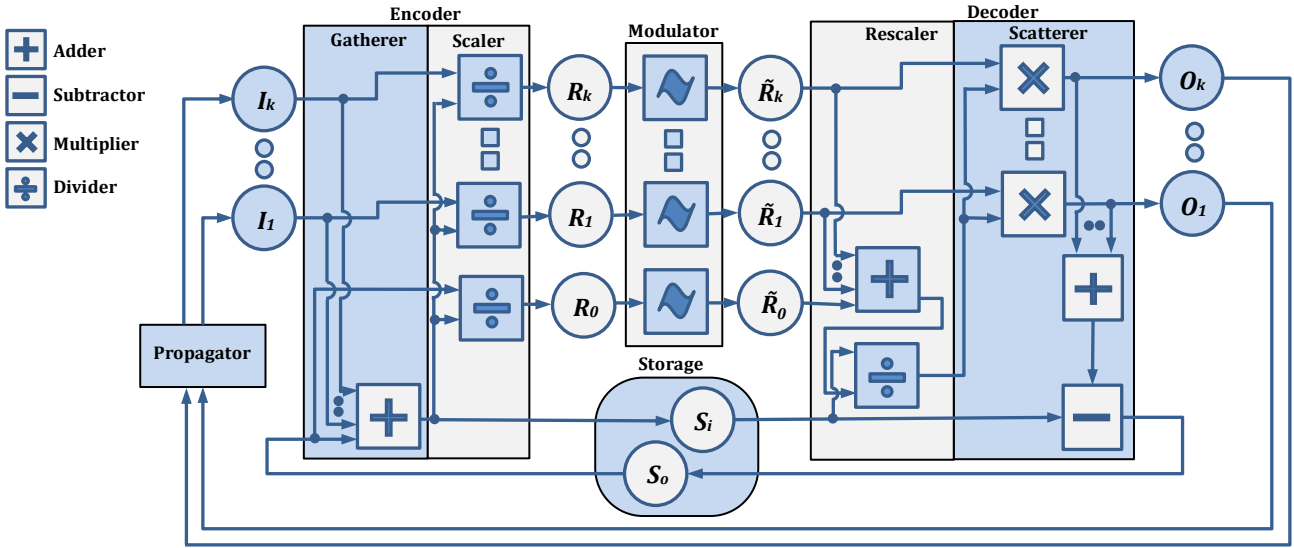

Fig.5 *Schematic diagram of the basic kinon model*

Since all operators are relational transformations or morphisms, the kinon model is very congenial to category theory and categorical system theory (Louie, 1983) in particular; thus the categorical meta-language and notation fit the structure and functioning of kinetic automata quite naturally. The categorical diagram in Fig.6 gives a compact representation of *the algorithm* of the basic kinon model and clearly shows the usage of storage in encoding and decoding and a pivotal role of the input storage $S_i$ during these steps.

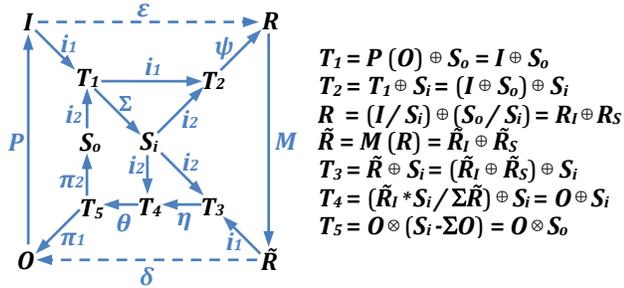

Fig.6 *Categorical diagram of the basic kinon model*

An isolated kinon, in which respective input and output buffers are looped, is possible but the collective behavior of kinons organized in a network is far more interesting. Formally, a kinon network is a *balanced digraph*, i.e. a directed graph in which the in-degree and out-degree of every vertex $v_i$, representing one kinon, are equal: $d^+(v_i) = d^-(v_i)$. A balanced digraph is said to be *regular* if all nodes have the same in-degree and out-degree. Zuse's net automaton and Pask's diffusion network, shown in Fig.1, exemplify regular and irregular kinon networks. Regular kinon networks, considered further, can be described by a node degree $d$ and lattice width $w$ (or a number of nodes $N$).

## Motivation for the model extension

It was shown in the previous paper that the relational approach and innate tunability of the model, i.e. the ability to be controlled by a smooth variation of one or more real-valued parameters, dramatically increase the complexity of its behaviour in comparison to continuous cellular automata. Nevertheless, the only tunable block in the basic model is the modulator, while other blocks are firmly hardwired. Encoding and decoding blocks, performing trivial but very important transformations, also can be made tunable via the elaboration of their circuitry, and these enhancements may have crucial consequences for the model's dynamics.

According to Robert Rosen, encoding is closely related to the problem of measurement and can be stated by the following propositions (Rosen, 1978):

• *The only meaningful physical events which occur in the world are represented by the evaluation of observables on states.*
• *Every observable can be regarded as a mapping (or encoding) from states to real numbers.*

This view is in line with the approach to measurement of the American psychophysicist Stanley Stevens who defined measurement as *"the assignment of numerals to objects according to a rule"* (Stevens, 1946). Initially, he identified four levels of measurement defined by groups of scale invariant mathematical transformations: *nominal*, *ordinal*, *interval* and *ratio*, but later added another scale type, *log-interval* (Stevens, 1959). However, this list is not complete and ratio is not the ultimate level of measurement. The ratio scale has one fixed point ('zero') and the choice of the value of 'one' is essentially arbitrary. An *absolute* level of measurement can be obtained if the value of 'one' is also fixed. The most apparent example of the absolute scale is probability, where the axioms fix the meaning of 'zero' and 'one' simultaneously.

The kinon model was derived from the LBM model based on statistical mechanics; nevertheless, it is fully deterministic. It equates the value of 'one' to the total amount of storage and inflow in the kinon, but it is invariant only during the current cycle; therefore a scaling step of encoding is related to a ratio scale. On the other hand, a scaling block transforms absolute (raw) input values corresponding to a nominal scale. The usage of other scales or evaluation methods in encoding may contribute to the overall complexity of the model's behaviour.

For that purpose, additional structural elements corresponding to electronic analog filters can be added to the encoder, which will process raw input values (observables) before scaling. Such filter can be treated as *a meter* evaluating input values via a nonlinear map, e.g. logarithmic or other function with a domain and codomain in $\mathcal{R}^+$. It will be a direct implementation of the Rosen's treatment of measurement as *'a mapping from states to real numbers'*, or formally $f: S \rightarrow \mathcal{R}$.

Another interesting option is the usage of a low-pass filter with memory known as *a leaky integrator*. A physical example of a leaky integrator is a bucket of water with a hole in the bottom. The rate of leakage is proportional to the depth of water depending on the difference between input and "leak" rates (hence the name). A very simple discrete time implementation of a leaky integrator is shown in Fig.7, where "*I*" stands for an input and "*Λ*" for a potential, which depends on the output in the previous time step and plays the role of water in a leaky bucket or short-term memory, which fades without reinforcement. This is akin to a moving average but does not require data buffering, which is very costly in computational and memory usage terms.

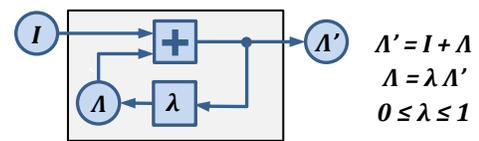

Fig.7 *Discrete time leaky integrator (λ-filter)*

Leaky integrators find their use in the neural and cognitive modeling (Graben et al, 2008) and the modeling of systems with anticipation property (Makarenko et al, 2007). An anticipatory system is a system that contains an internal predictive model of itself and its environment, which allows it to change the current instant state in accordance with the model's predictions pertaining to a later instant (Rosen, 1985).

A leaky integrator is, perhaps, the simplest model capable of predicting the future state of a system and the easiest way to introduce a field approach in the kinon model. It can be implemented by linking additional variables to input buffers and storage, which will represent the channel potentials and play the role of local curvature or space. In this case, potentials will be influenced by input flow and, in their turn, influence output flow, representing matter. This is analogous to the famous quote by John Wheeler: *"Matter tells space how to curve and curved space tells matter where to move"*.

The usage of cut-off (threshold) filters for the elimination of unwanted marginal values or simulation of surface tension would be also beneficial for the overall model nonlinearity.

## Extended model

A gathering block of the encoder was augmented by the embedding of $\lambda$-filters in all input and storage channels. They are leaky integrators with a common tunable real-valued control parameter $\lambda$ in a unit range, representing a memory "leak" rate. A scaling block was enhanced by the incorporating of $\psi$-filters which transform input absolute values via a nonlinear function. The nonlinearity is vital here because modulation is scale invariant. This operation is similar to gamma-correction used in image processing for the adjustment of pixel intensity values according to human visual perception. The letter $\psi$ is frequently used as a symbol of psychology and perception, so it was chosen for the name of the filter. The introduced filters are shown in Fig.8 and highlighted by a peach color.

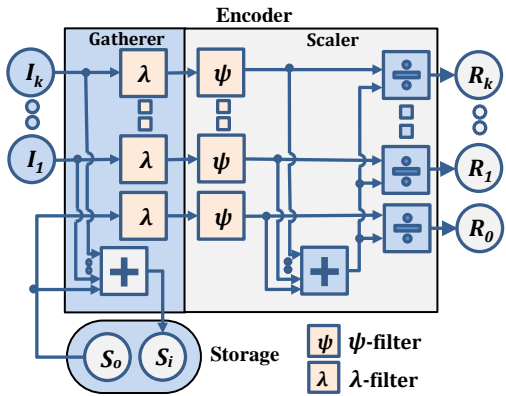

Fig.8 *Schematic of the tunable encoder*

Similarly, a decoding module, consisting of the rescaler and scatterer, was enhanced by adding two new kinds of analog filters shown in Fig.9. $\theta$-filters truncate rescaled absolute output values below a threshold before scattering them into output buffers. They are tuned by a common real-valued control parameter $\theta$ in the range $[0, \Theta]$, where $\Theta$ is much less than the total quantity of the network. It can be treated as the simplistic simulation of fluid cohesion due to surface tension.

Another novel filter, shown in Fig.10 in more detail, is analogous to $\lambda$-filter, i.e. a leaky integrator, but is a little different. Similar to a conventional leaky integrator, it integrates values obtained on different time steps, but in this case, both time steps take place during the same cycle. A tunable parameter $\eta$ defines here not a storage 'leak' rate but a share of the input storage value not participating in distribution (decoding) and remaining in the storage. In medicine, a small passage which allows movement of fluid from one part of the body to another is called a shunt, so a shunting integrator is a more proper name for such a filter. It can be treated as the simplistic simulation of fluid viscosity, a quantitative measure of fluid resistance to flow drag, generally denoted by $\eta$. Its influence is similar to damping in mechanical systems, although it is not based on energy loss, so a damper might be another "telling" name for the $\eta$-filter.

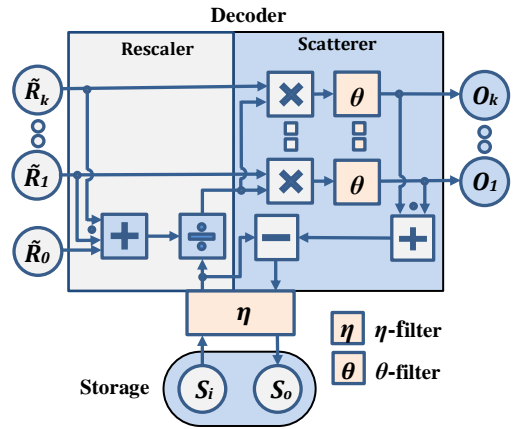

Fig.9 *Schematic of the tunable decoder*

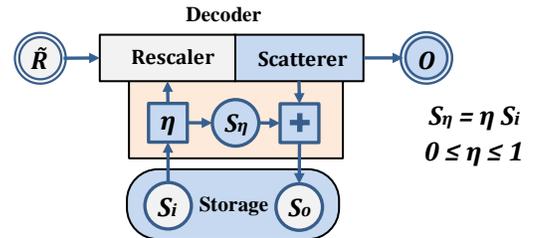

Fig.10 *Shunting integrator ($\eta$-filter)*

All introduced filters, denoted as $f_\lambda$, $f_\psi$, $f_\eta$ and $f_\theta$, with new variables corresponding to channel potentials ($\Lambda$, $\Lambda'$), their measurements (percepts) ($\Psi$, $S_\psi$) and fractions of input storage ($S_\eta$, $S_\delta$) can be succinctly represented by the following categorical diagram of the extended kinon model [Fig.11], which represents *the full algorithm* of the kinon model. All alterations and additions against the categorical diagram of the basic kinon model [Fig.6] are shown in a red color.

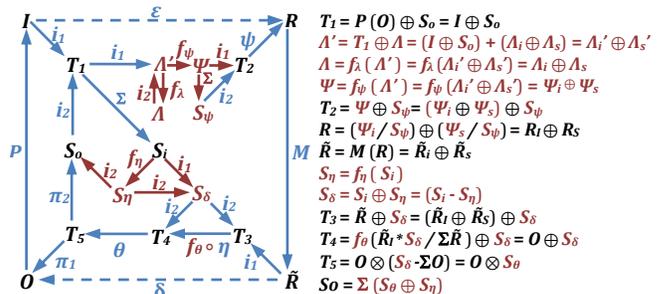

Fig.11 *Categorical diagram of the extended kinon model*

# Results

All results presented further are obtained using the simplest kinetic map: $y = Max[0, (1-\kappa x)]$. It was demonstrated in the previous paper that this map, despite its simplicity, exhibits highly complex behaviour with phase transitions. It has a single real-valued control parameter $\kappa$, which substantially simplifies the description of the parameter space and clarifies the influence of other parameters on the morphology of generated patterns.

Since this paper is aimed to show the applicability of the kinon model to morphogenesis, the evolution of the kinon network always starts from a 'singularity' state, corresponding to a zygote state in biological morphogenesis. It means that only one kinon has a non-zero storage equal to the total quantity (energy) of the network. This value is set to 20 000 for a 200 x 200 square grid, which is equivalent to 0.5 average value per a kinon visualized as a grey color in a greyscale image. It never changes during the evolution of the kinon network because quantity conservation is a staple feature of the model.

Apart from a kinetic map parameter $\kappa$ in the basic model, the extended model has a set of additional parameters $\{\lambda, \psi, \eta, \theta\}$ controlling the introduced filters. The paremeters $\lambda, \eta, \theta$ are real-valued non-negative numbers equal to a zero by default, while $\psi$ is a function with characteristic parameters. The latter parameter is equal to the identity map *id: y=x* if another is not specified.

It is not surprising that the most significant parameter is $\lambda$ because it controls internal potentials relating to kinon 'memory' or 'anticipation' property. Just two parameters, $\kappa$ and $\lambda=1$, are sufficient for the appearance of circular or square waves not observed in the basic model. The increment of the parameter $\kappa$ results in the change of the shape and length of the wave and the speed of its final fission into four solitons [Fig.12]:

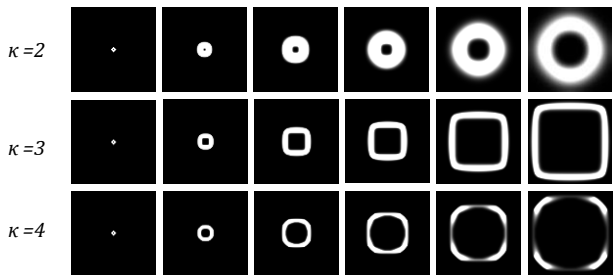

Fig.12 *Two-dimensional kinetic waves* ($\lambda=1$)

The adjustment of scattering $\theta$-filters, which can be related to the change of surface tension, induces the appearance and development of manifold shapes. Using the same parameters $\kappa$ and $\lambda$ as above and the parameter $\theta=2$, one can obtain dynamical patterns reminding the embryonic development of a 'tetrapod star fish' or a growth of a four-fold crystal [Fig.13]. The increments of the parameter $\kappa$ result here in the dramatic changes of the final morphology of the creature.

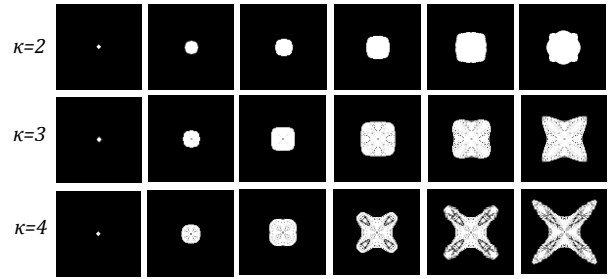

Fig.13 *Kinetic morphogenesis* ($\lambda=1$ $\theta=2$)

In all these examples, development starts from a single cell and goes through the stages which can be related to the oocyte, blastula, gastrulation and organogenesis stages in real biological morphogenesis. Due to the parameter $\theta$, the developmental process finally stops at the state of a stable kinetic equilibrium corresponding to a mature full-fledged stage or stasis, which will be examined in the next paragraph in more detail.

For brevity, further demonstrations will show only the final stable state accompanied by a drawing of contour lines (isolines) after every 20 time steps (cycles). The results shown in Fig.14 demonstrate the influence of $\eta$-filter on the size and shape. According to a 'damping' nature of the $\eta$-filter, the final shape becomes more and more compact after the increment of this parameter.

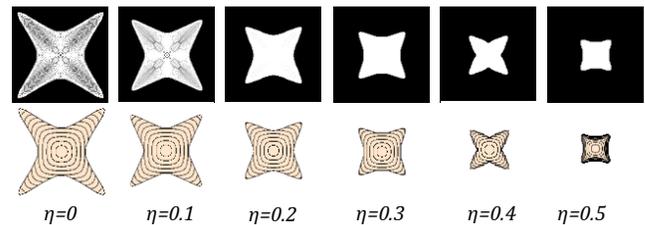

Fig.14 *η-patterns* ($\kappa=3$ $\lambda=0.5$ $\theta=1$)

The influence of $\psi$-filter, which is closely related to the earlier discussed measurement problem and perception, cannot be so easily estimated. Similar to a kinetic map, it is a *functor* rather than a function, i.e. it maps functions but not values. The outcomes of the application of some $\psi$-filters under the same other parameters are shown in Fig.15.

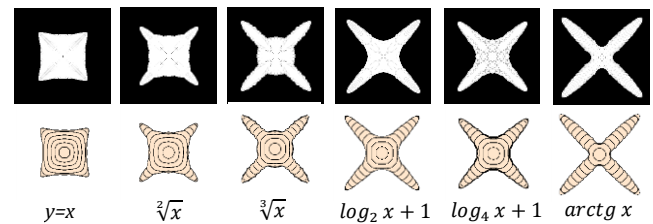

Fig.15 *ψ-patterns* ($\kappa=3$ $\lambda=1$ $\theta=1.5$ $\eta=0.1$)

It is evident that this filter affects not only the size but also the morphology of the resultant shape. Due to increasing nonlinearity of the function, the growing nucleus develops a branching structure that becomes more stretched and pointed.

The shown patterns were obtained using a square grid with four nearest orthogonal neighbors, i.e. a four-degree (*d4*) network. They have a distinct four-fold symmetry imposed by the underlying grid but preferential directions are aligned along the diagonals of the axes. This counter-intuitive phenomenon is related to the kinetic fission demonstrated in Fig.12, but is yet unexplained and needs further investigation.

The kinon model is topology invariant and allows arbitrary network structure including meshes and random networks. A *d4* network was chosen only for the ease of computation and visualization. A square grid with eight nearest neighbors (*d8*) requires extra computation per a cycle but is also readily visualized by a greyscale image.

Fig.16 demonstrates *ψ*-patterns in a *d8* network. The main difference from the previous experiment is the change of the preferred growth directions from diagonal to orthogonal. Besides, some new features of the body plan have appeared. Branches now may have not only pointed tips but also cleft and undulating ones.

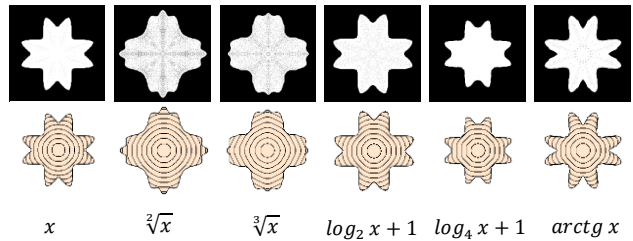

Fig.16 *d8 ψ-patterns (κ=8 λ=0.8 θ=0.3 η=0.4)*

The increased network connectivity makes kinon dynamics less predictable and parameter changes usually affect both the size and shape. Fig.17 demonstrates the susceptibility of the shape to minor variations of a single parameter, which indicates possible amenability of the extended kinon model to genetic and other evolutionary algorithms.

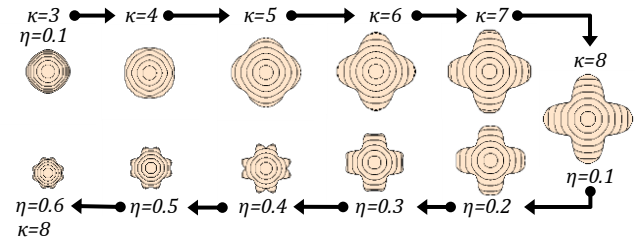

Fig.17 *Kinetic "metamorphosis" (d=8 λ=1 θ=0.6)*

The collections of some characteristic kinon morphogenetic patterns obtained using *d4* and *d8* grids are shown in Fig.18.

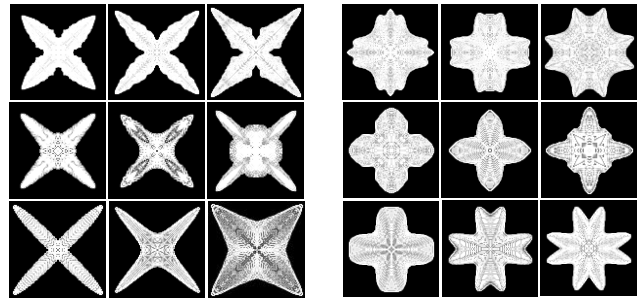

Fig.18 *Panopticon of d4 (left) and d8 (right) kinon creatures*

Speaking about a network structure, it is essential to define its boundary conditions. In practice, one cannot deal with an infinite lattice, so a common approach is to assume periodic (or cyclic) boundary conditions, i.e. to embed a lattice in a torus. However for morphogenetic studies, a better solution is the correct choice of the grid size and model parameters sufficient for a full-fledged state. Another option is the imposing of artificial boundaries (borders) on the network for the study of bounded growth. Due to topological invariance, the kinon model allows simple implementation of boundary conditions by the permanent elimination of respective links in the boundary kinons. The examples of bounded growth with different borders are shown in Fig.19.

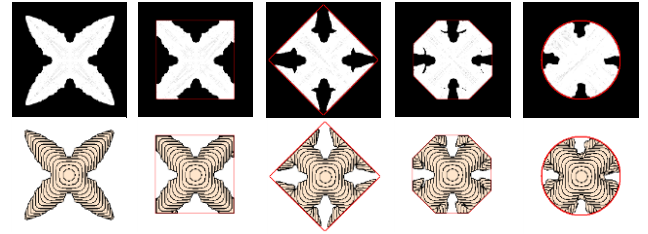

Fig.19 *Bounded growth (d=4 κ=6 λ=0.8 θ=0.4 η=0.5)*

## Macrodynamic analysis

The total quantity of the kinetic network, which can be considered as energy or mass, is always the same because the model is conservative by definition. However, the total amount of all kinon output buffers (related to external or kinetic energy) and the total amount of kinon storage (related to internal or potential energy) interchangeably fluctuate. This feature can be used in the quantitative analysis of the kinon network macrodynamics.

The simplest macrodynamic index termed as an *exchange rate* $K_e$, is the ratio of the total value of all kinon output buffers to the total quantity of the network. It also can be interpreted as the ratio of the kinetic energy to the total energy of the network and is calculated as follows [Eq.1]:

$$K_e = \frac{\sum_{i=1}^{n}\sum_{j=1}^{k} O_{ij}}{\Omega} \in [0, 1], \quad (1)$$

where: $n$ - the number of kinons in the network;
$k$ - the number of neighbors of the $i^{th}$-kinon;
$O_{ij}$ - the $j^{th}$-output buffer value of the $i^{th}$-kinon;
$\Omega$ - the total quantity of the network.

Another macrodynamic index, also having a unit range and termed as *a turnover rate* $K_t$, is a half of the ratio of the absolute change of all kinon buffers during the current cycle to the total quantity of the network [Eq.2]:

$$K_t = \frac{\sum_{i=1}^{n}(|\Delta S_i| + \sum_{j=1}^{k}|\Delta E_{ij}|)}{2\Omega} \in [0, 1], \quad (2)$$

where: $\Delta E_{ij}$ - the difference of respective exchange buffers ($I_{ij} - O_{ij}$) of the $i^{th}$-kinon;
$\Delta S_i$ - the change of the $i^{th}$-kinon storage.

Even such simple macrodynamic indices can tell a lot about the current state and dynamics of the whole network [Fig.20]:

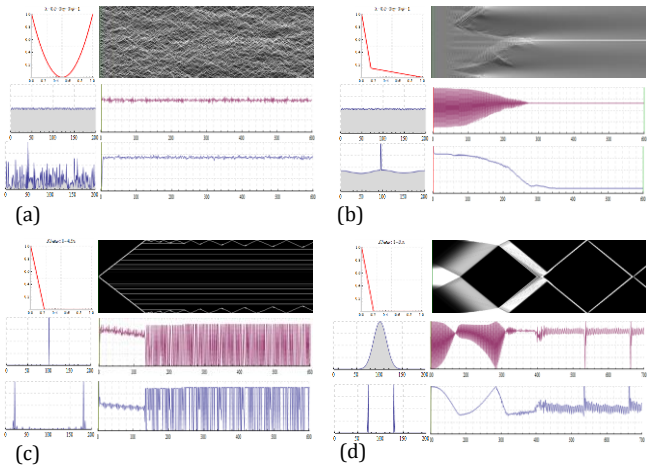

(a) (b) (c) (d)

Fig.20 *Macrodynamics of one-dimensional kinon networks*

The first rows in these examples show the kinetic maps and visual images of the dynamics of one-dimensional kinon networks, which initial and ending states are represented by histograms in the left below. The plots below images represent the dynamics of exchange rates *Ke* and turnover rates *Kt* depicted in red and blue colors respectively.

Fig.20a shows the dynamics in a chaotic state, while fig.20b demonstrates the transition from a nearly equilibrium state to a stable non-uniform state. Both examples start from the same initial state but show totally different behaviour. Despite the final visual stasis in the latter figure, both indices attain a constant non-zero value. It means that the kinon network reached a coherent dynamic state, which can be related to Nash equilibrium in game theory. Fig.20c displays an almost still picture after the collision of solitons, sharply contrasting to the behaviour of both indices, oscillating in a nearly full range. Fig.20d illustrates the behaviour of these indices during the fission of the bell-shaped expanding wave in two solitons and their subsequent collisions.

Macrodynamics of some characteristic morphogenetic patterns is represented in Fig.21. The green line marks the beginning of stasis and the orange line marks the time step when dendrite tips hit the border.

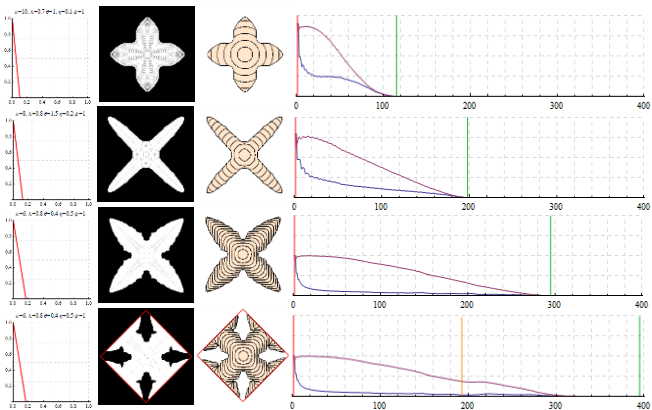

Fig.21 *Kinetic macrodynamics of morphogenesis*

These indices supplement visual representation of the kinon network dynamics and are indispensable in cases when visualization is intractable or the automation of parameter space exploration is needed, e.g. for the search of viable stable shapes. All demonstrated morphogenetic examples come to a matured full-fledged stable state (stasis) in which both macrodynamic indices reach a zero value.

## Discussion

The shown results demonstrate that the extended model, employing only trivial math, is capable of producing complex patterns, some of which resemble crystal dendrites. Dendritic crystal growth is very common and may be illustrated by snowflake formation and frost patterns on a window. In metallurgy, a dendrite is a characteristic tree-like structure of crystals growing during molten metal solidification. This dendritic growth has large consequences in regard to material properties. That is why much research has been devoted to the simulation of crystal growth, and one of the most employed approaches is a phase-field model. Despite the enormous progress in computational terms, a phase-field model, based on the Ginzburg-Landau equation, still requires quite hard computation. The kinon model was not tailored to simulate dendritic growth in melts; nevertheless, it captures this phenomenon rather closely and generates very similar shapes [Barrett et al, 2014].

Pattern generation and morphogenesis are not restricted to physical and biological systems and may be realized in many other contexts. In Robert Rosen's words: *"Morphogenesis, in the widest sense, is the generation of pattern and form in a population of interacting elements"*. He showed that morphogenetic techniques can be also applied to human settlement patterns and the emergence of cultural differentiation in a human population (Rosen, 1979). He was, possibly, the first to assert that active transport of materials against their diffusion gradients is ubiquitous in biology and to show that the coupling of reactions and diffusion can result in surprising effects, such as the accumulation of population against density gradients. The coupling of diffusion with reaction can play the role of specific "pumps" moving populations uphill their gradients.

Moreover, Rosen envisaged the possibility of asymptotically stable non-uniform states with just one morphogen, rather than two as in Turing's original paper, which was recently rediscovered by the author using a different approach. Turing implicitly assumed that each cell was homogeneous and isotropic with respect to diffusion in all directions. Rosen showed that it is perfectly possible to construct a simpler system by removing the assumption of diffusional isotropy (Rosen, 1981).

The kinon model proved to be able to produce complex shapes using very simple local interactions by the exchange of real-valued numbers, which can be regarded as the quantities of the elusive long-sought chemical signal, which Turing called a morphogen. A large collective of kinons embodied in a robot swarm would gradually aggregate into dendritic or more complex structures by the exchange of such morphogen with the nearest neighbors.

Contrary to another kinetic critical phenomenon called diffusion-limited aggregation (DLA) (Witten & Sander, 1981), this model generates totally deterministic rather than random shapes. The kinon model does not need a global clock, but does require local synchronization. Unlike static regular kinon networks, considered in this paper, a swarm network topology is irregular in a general case and node neighborhoods are variable. In order to cope with these circumstances, the necessary mechanisms for local interaction and synchronization must be added to the propagator.

All figures demonstrating kinon network states and dynamics [Fig.12-21] were obtained using a framework called *KinonLab*, developed by the author. It is implemented in Wolfram *Mathematica*® and currently supports one-dimensional (*d2*) and two-dimensional regular networks with von Neumann (*d4*) and Moore (*d8*) neighborhood [Fig.22].

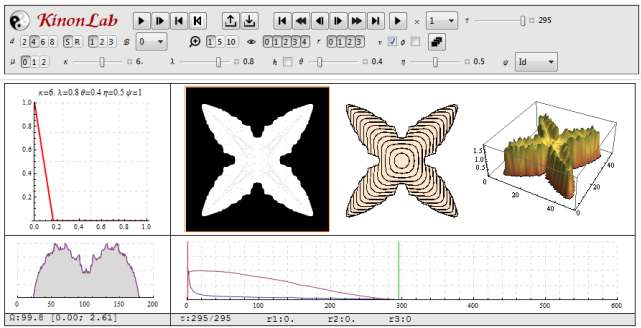

Fig.22 *KinonLab screenshot*

The main directions of the ongoing improvement include the support of two-dimensional hexagonal (*d6*) and three-dimensional regular grids, and also the development of tools for micro-dynamic and comparative analysis. The future developments of the framework will be aimed at the study of *multi-component* and *arbitrary topology* kinon networks.

## Conclusion

The presented here extended kinon model demonstrates its high potential for structural elaboration and the need for deeper exploration and validation. The given formal definition clarifies some ambiguities of the previous description and provides a common language and terminology to facilitate its further development. The schematic and categorical diagrams of the kinetic automaton not only illustrate its operational structure ('modus operandi') but also show the directions of the possible implementation of the model with analog circuits. The proposed macrodynamic indices provide the expressive measures for the quantitative analysis of kinon network dynamics and parameter space exploration.

The main aim of this paper is to demonstrate the applicability of the model to the problem of morphogenesis. The shown results confirm that the extended model is capable of producing spatio-temporal patterns pertaining to morphogenesis in real physical and biological systems. In respect to Artificial Life research, the most promising directions of the kinon model application are rapidly emerging fields of morphogenetic engineering (Doursat et al, 2013) and swarm robotics (Brambilla et al, 2013).

## Acknowledgments

I would like to thank Dr. Aloisius H. Louie for his expert advice and warm encouragement during the elaboration of the categorical kinon diagrams.